\documentstyle[twoside,fleqn,espcrc2]{article}

\newcommand{\AmS}{{\protect\the\textfont2
  A\kern-.1667em\lower.5ex\hbox{M}\kern-.125emS}}

\title{
The (dual) Meissner effect in SU(2) and SU(3) QCD
\thanks{Presented by Y. Matsubara.}}

\author{
Yoshimi Matsubara
 \address{ Nanao Junior College, Nanao, Ishikawa 926, Japan}
, Shinji Ejiri
 \address{ Department of Physics, Kanazawa University,
         Kanazawa 920-11, Japan}
and Tsuneo Suzuki
\addtocounter{address}{-1}
\addressmark
}

\begin{document}

\begin{abstract}
After making an abelian projection in the maximally abelian gauge, we measure 
the distribution of abelian electric flux and monopole currents around
an abelian Wilson loop in $SU(2)$ and $SU(3)$ 
QCD.  The (dual) Meissner effect is observed clearly. The vacua in the 
confinement phases of $SU(2)$ and $SU(3)$ are both at around the border between
type-1 and type-2 (dual) superconductor. 
\end{abstract}

\maketitle

\section{Introduction}
 It is important to understand the mechanism of quark confinement in QCD.
 We expect that color electric flux is squeezed by the dual Meissner effect
 due to color magnetic monopole condensation\cite{thooft,mandel}. 
It is well known that such a
situation occurs in compact U(1) lattice gauge theory(LGT)\cite{bank}. 
In QCD, 'tHooft 
conjectured that the role is played by the monopoles in abelian
projected QCD which can be regarded as an abelian theory with abelian
charges and magnetic monopoles\cite{abelpro}. 
We choose the maximally abelian (MA) gauge \cite{kron}
in which diagonal components of all link variables are maximized 
as much as possible by the gauge transformation.
Gauge-fixed link variables are decomposed into a product of two matrices
$\tilde{U}(s,\mu)=V(s)U(s,\mu)V^{\dag}(s+\mu)=c(s,\mu)u(s,\mu),$
 where $V(s)$, $c(s,\mu)$ and  $u(s,\mu)$ are 
a gauge-fixing matrix, an off-diagonal and a diagonal matrix, respectively. 

 Monte Carlo simulations in this gauge have shown interesting 
results, 
i.e., abelian dominance \cite{yotsu,hio,suzu}
and interesting monopole behaviors \cite{kron,hio,suzu,kitahara}.
These results show that the monopoles play an important role 
in QCD confinement, but evidence of the monopole condensation 
in the confinement phase is not yet obtained. 

In this report,
we show the (dual) Meissner effect occurs in the presence of external charges 
also in QCD\cite{hay,cea}.
In the continuum, the assumptions of the abelian dominance and the monopole 
condensation lead us 
to a dual form of the abelian Ginzburg-Landau (G-L) theory\cite{suzuki}.
So we study the G-L type equations between the abelian electric 
field and the magnetic monopole currents in the presence of a static
quark-antiquark pair in SU(2) and SU(3), following the method 
developed by Singh et al.\cite{hay}.
Our purpose is to apply the method
to SU(3) LGT as well as SU(2) and to investigate scaling behaviors of the 
penetration length $\lambda$ and the coherence length $\xi$.

\section{
Simulations and measurements
}
 We used the Wilson action and simulations are performed on 
          $ 16^4$ lattice in SU(2) ($\beta=2.4,2.45,2.5$ and $2.6$) 
and 
       on    $10^4$ lattice in SU(3) ($\beta=5.6,5.7,5.8$ and $5.9$).
Measurements are done every 30 sweeps and 
we adopted totally 400 configurations in SU(2) and 500 configurations in SU(3).
For thermalization, we discarded inital 1500 sweeps.
Abelian link variables $u(s,\mu)$
are extracted after the gauge-fixing is done in the MA gauge.
The average of an observable $O(u)$ is computed as 
\begin{equation}
  <O(u)> = \frac{{\rm Tr}e^{-S}W(I,I)O(u)}{{\rm Tr}e^{-S}W(I,I)},
\end{equation}
where $W(I,I)$ is an abelian Wilson loop. We used $3 \times 3$ and 
$5 \times 5$ in SU(2) and $3 \times 3$ in SU(3).
 The electric field operator is defined by 
\begin{equation}
           E_i(s) ={\rm Im}(u_p (s,i,4)),
\end{equation}
where $u_p (s,\mu,\nu)$ is an abelian plaquette operator.          
 Monopole currents are defined following  DeGrand-Toussaint\cite{degrand}.

We investigate the dual G-L equations\cite{hay}: 
\begin{equation}
\vec{E}(r)  = \vec{\nabla} \times \vec{A}(r),
\end{equation}
\begin{equation}
\vec{J}_m (r) =  f(r)^2(\vec{A}(r)-\Phi\vec{\nabla}\alpha/(2\pi))
/\lambda^2,
\end{equation}
\begin{eqnarray}
-\xi^2 \nabla^2f(r)+\xi^2 (\vec{\nabla}\alpha-2\pi\vec{A}(r)/\Phi)^2 f(r)\\ 
\nonumber-f(r)+f(r)^3=0. \label{fx}
\end{eqnarray}
The order parameter $f(r)$ is given by the 
following approximate solution of the G-L equation
\begin{equation}
        f(r) =\tanh(\nu r/\xi), \label{fr}  
\end{equation}
where we adopted $ \nu =1.0$. 
Then we get 
the following extended London equation:
\begin{equation}
 \vec{E}(r) - \lambda^2 \vec{\nabla} \times ( \vec{J}_m(r) /f(r)^2 )
=n\Phi\delta (r),
 \label{er}\\
\end{equation}
where $n$ is an integer and $\Phi$ is the quantized electric flux 
at the origin $r=0$.

\section{
Analyses
}

 To reproduce the  $-\vec{\nabla} \times \vec{J}_m$data, 
we choose the following 
function for the azimuthal component 
of the monopole current:
\begin{equation}
           J_m (r)=a_1 (r+a_2 r^2+a_3 r^3)\exp (-a_4 r), \label{jm} 
\end{equation}
where $a_1 , a_2 , a_3$ and $a_4$ are parameters. 
The form ensures that $J_m (r)$ 
vanishes linearly at $r=0$ and exponentialy as $r \rightarrow \infty$.
Then, substituting the current into Eq.\ref{er},  we 
examine the electric field data 
except the $r=0$ point and determine both the penetration and the coherence 
lengths.              
The typical data of the curl of the monopole currents and the electric field
are plotted in Fig.1 ($SU(3)$). 
 If the extreme type-2 dual superconductor is the case,  $f=1$.
 Then, the data with
 $-\vec{\nabla} \times \vec{J}_m>0$ and $E>0$ can not satisfy Eq.\ref{er}.
Hence, there must be a region where $ f \neq 1 $, i.e., 
we need both $\lambda$ and $\xi$
 also in the SU(3) case.
 The values of the both lengths obtained are plotted in Fig.2.
 The fluxoid $\Phi$ evaluated from the fluxoid relation is consistent with
the net flux through the entire lattice in SU(2) and SU(3).

The G-L parameter $ \kappa = \lambda / \xi $ are shown in Fig.3.
The values are close to the borderline of the type-1 and
 type-2 dual superconductor.
It seems that the data of the large Wilson loop
 show the scaling behaviors.

\input epsf

\begin{figure}[htb]
\vspace{-20mm}
\epsfxsize=0.5\textwidth
\begin{center}
\leavevmode
\epsfbox{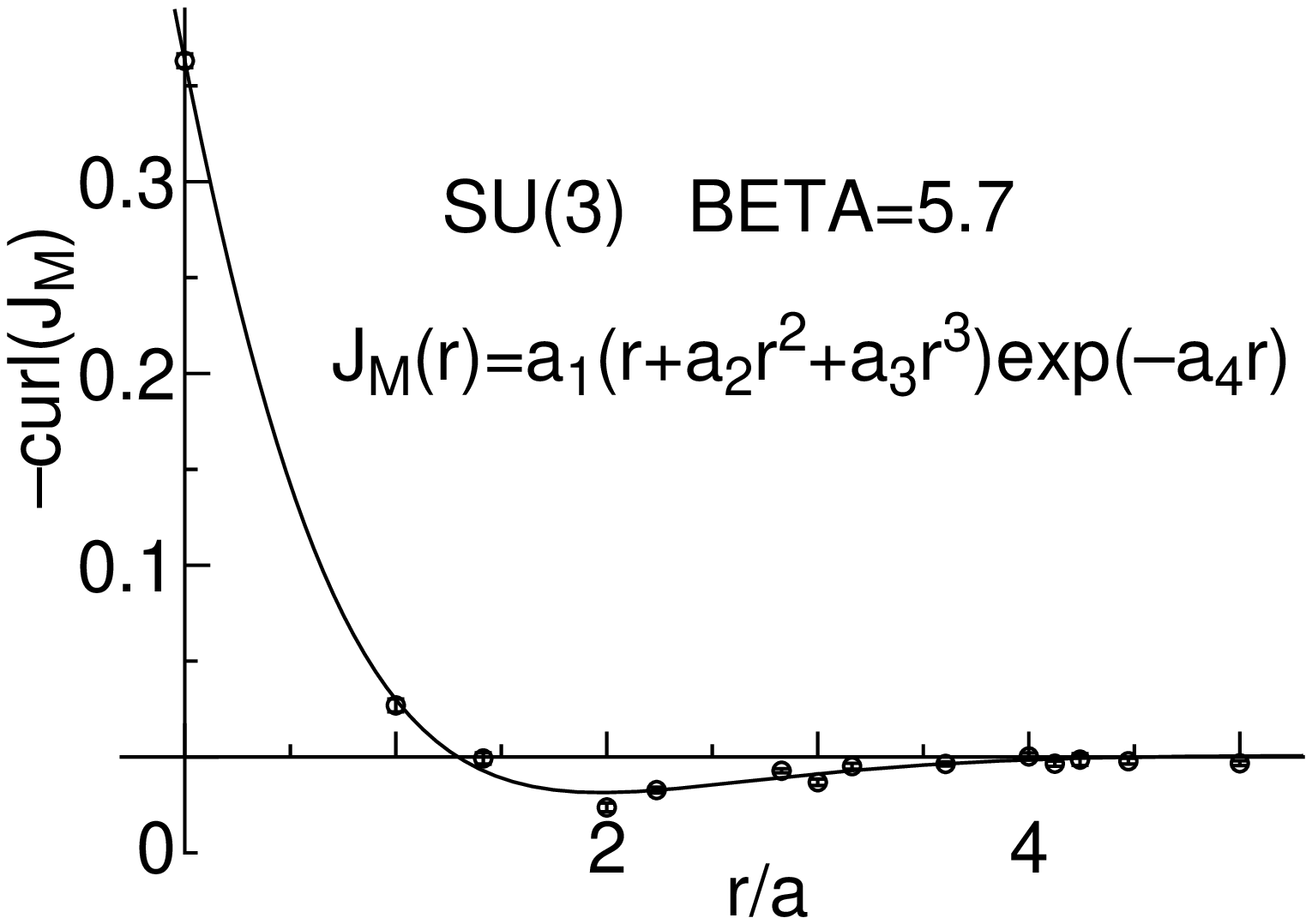}
\end{center}
\vspace{-50mm}
\epsfxsize=0.5\textwidth
\begin{center}
\leavevmode
\epsfbox{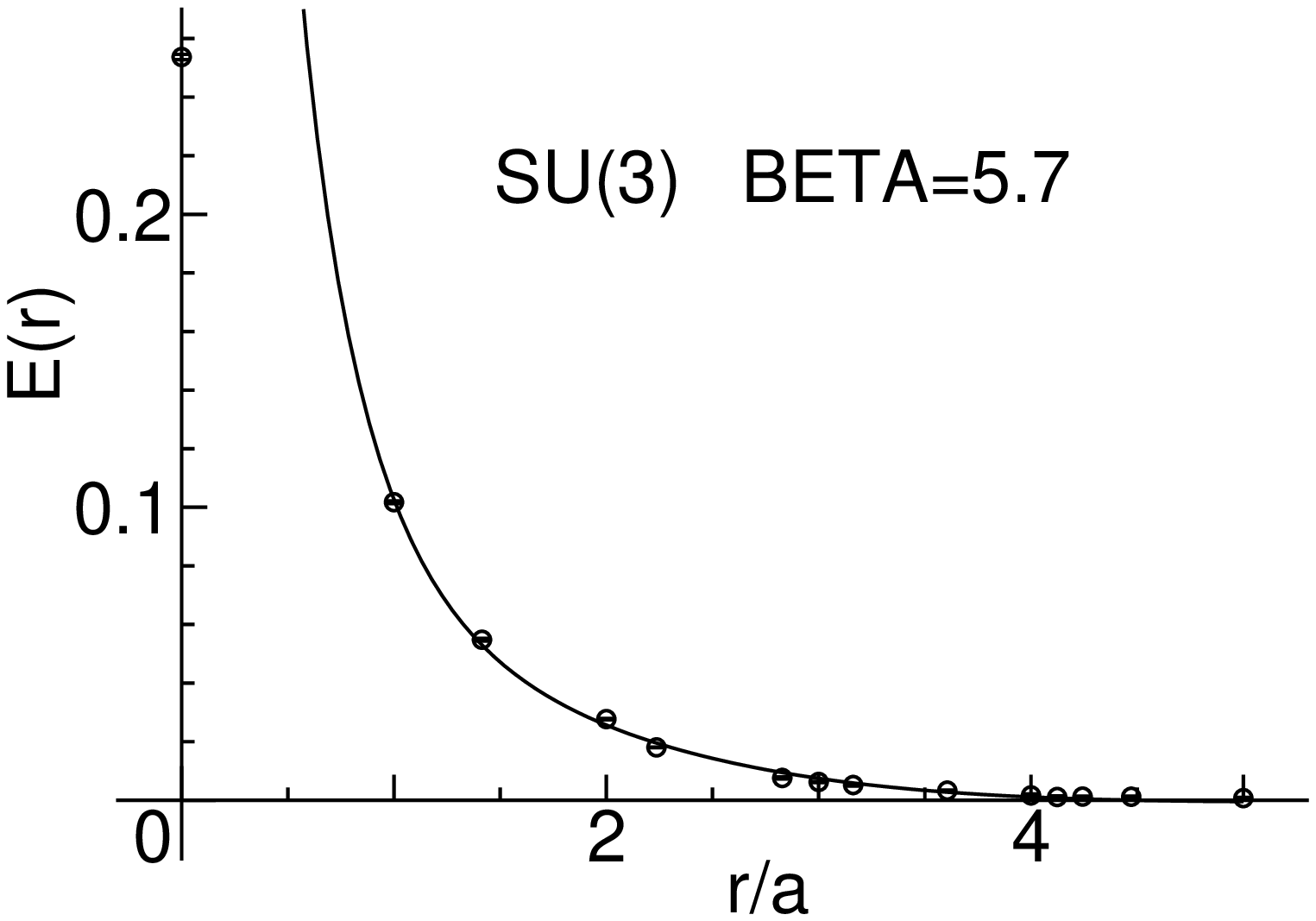}
\end{center}
\vspace{-43mm}
\caption{
The curl of the monopole current (upper) and the electric field (lower) 
in $SU(3)$. 
}
\vspace{-5mm}
\label{fig:curl3}
\end{figure}

\begin{figure}[tb]
\vspace{-20mm}
\epsfxsize=0.5\textwidth
\begin{center}
\leavevmode
\epsfbox{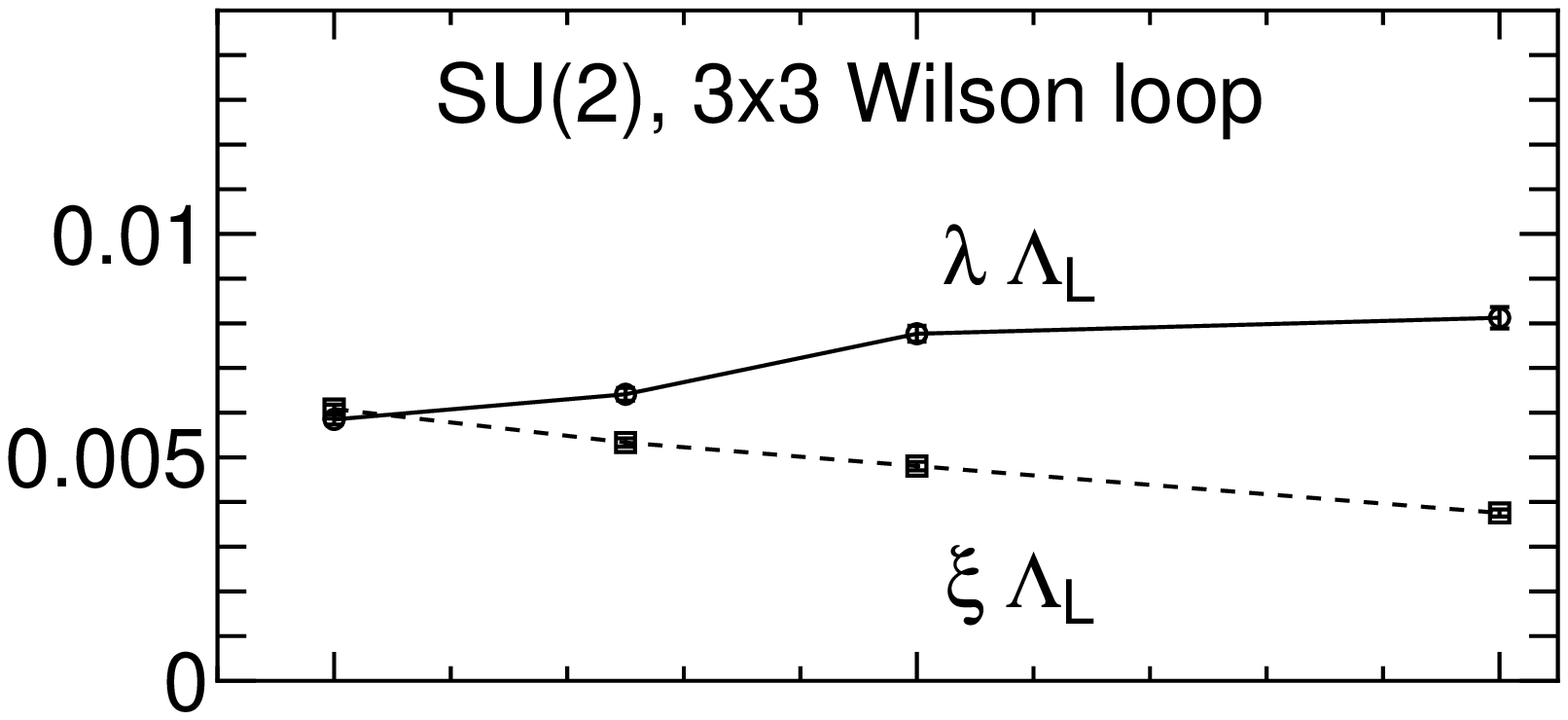}
\end{center}
\vspace{-64mm}
\epsfxsize=0.5\textwidth
\begin{center}
\leavevmode
\epsfbox{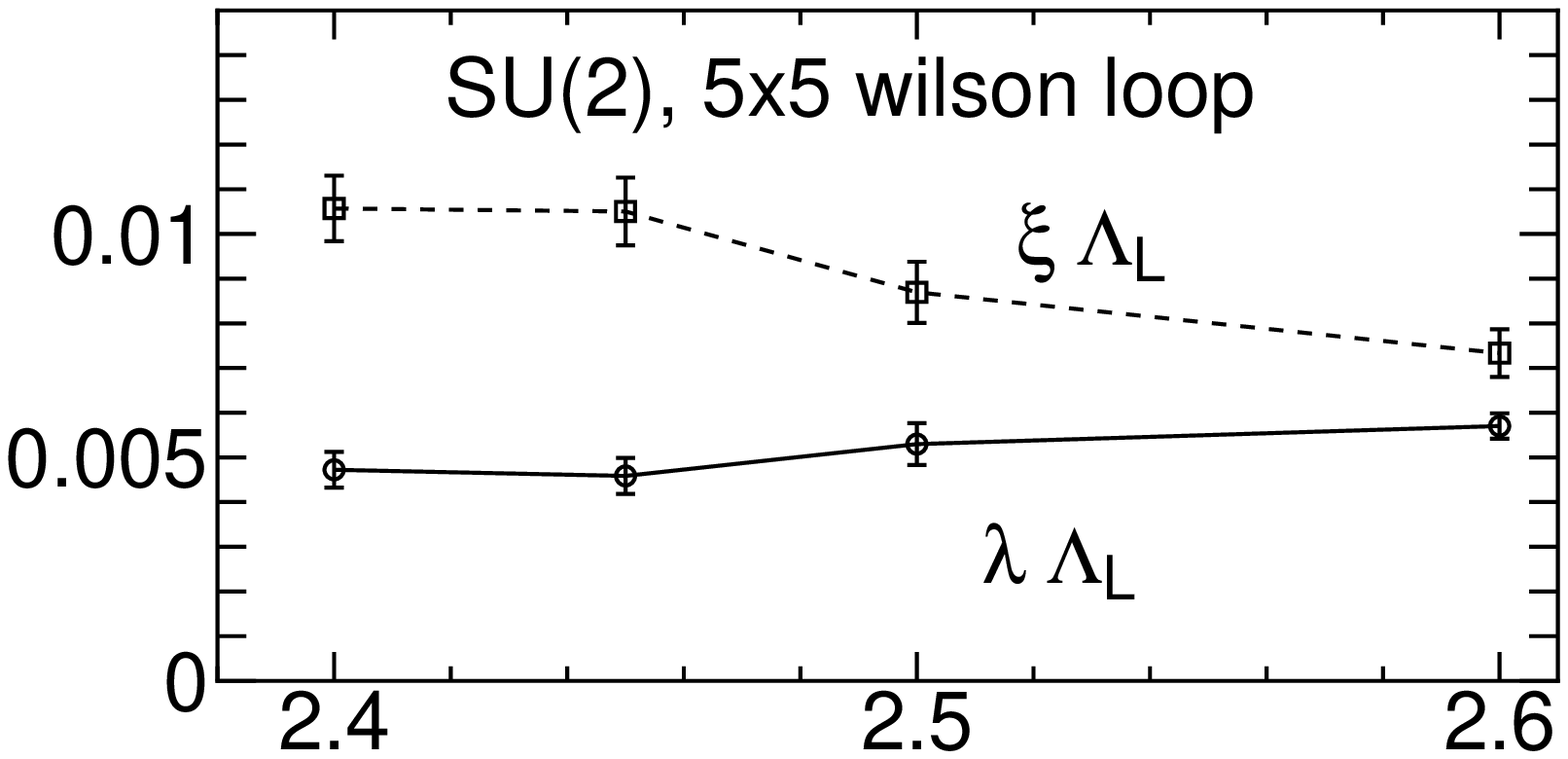}
\end{center}
\vspace{-61mm}
\epsfxsize=0.5\textwidth
\begin{center}
\leavevmode
\epsfbox{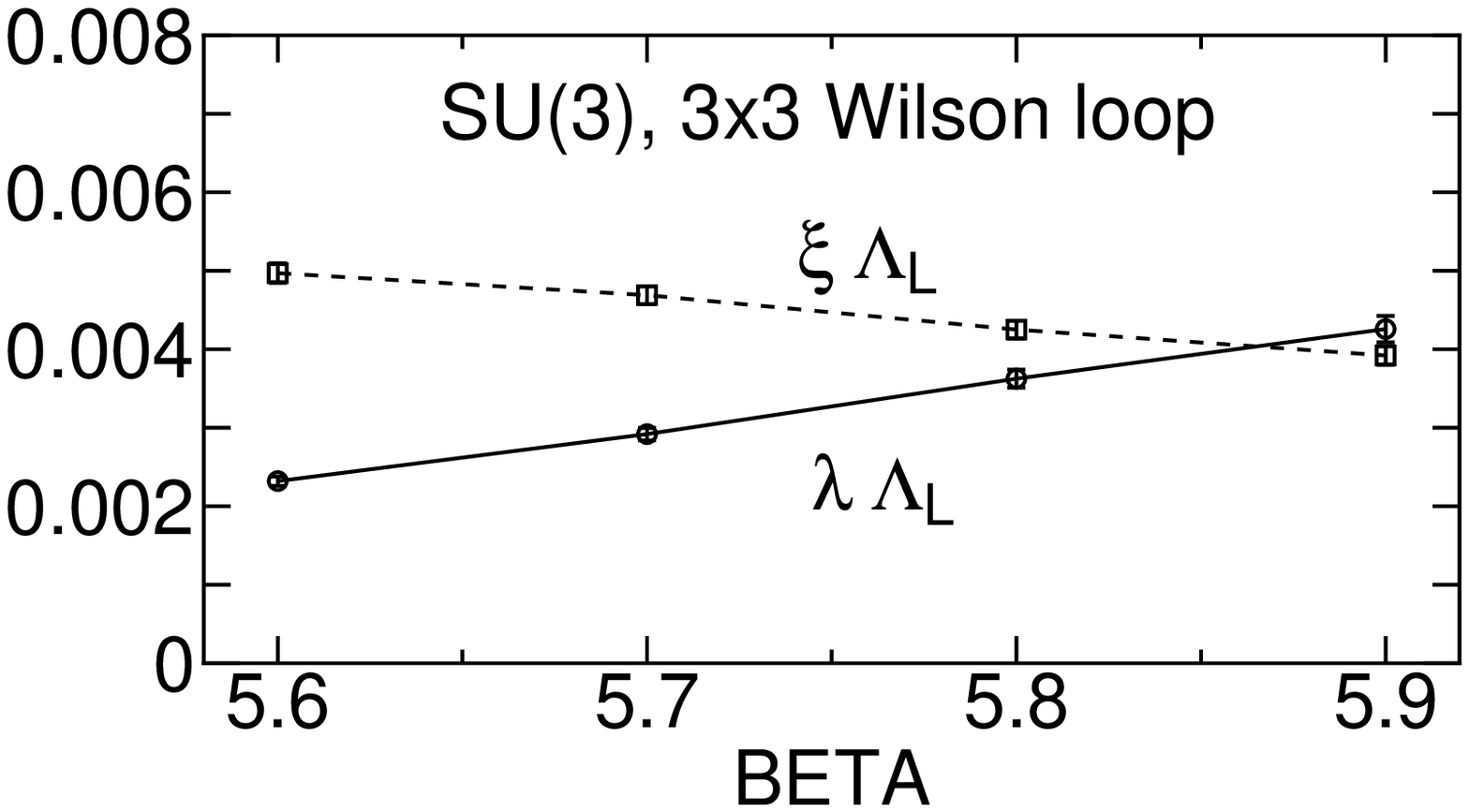}
\end{center}
\vspace{-50mm}
\caption{
The penetration and the coherence lengths versus $\beta$ in $SU(2)$ and $SU(3)$.}
\vspace{-5mm}
\label{fig:scale1}
\end{figure}

\begin{figure}[tb]
\vspace{-20mm}
\epsfxsize=0.5\textwidth
\begin{center}
\leavevmode
\epsfbox{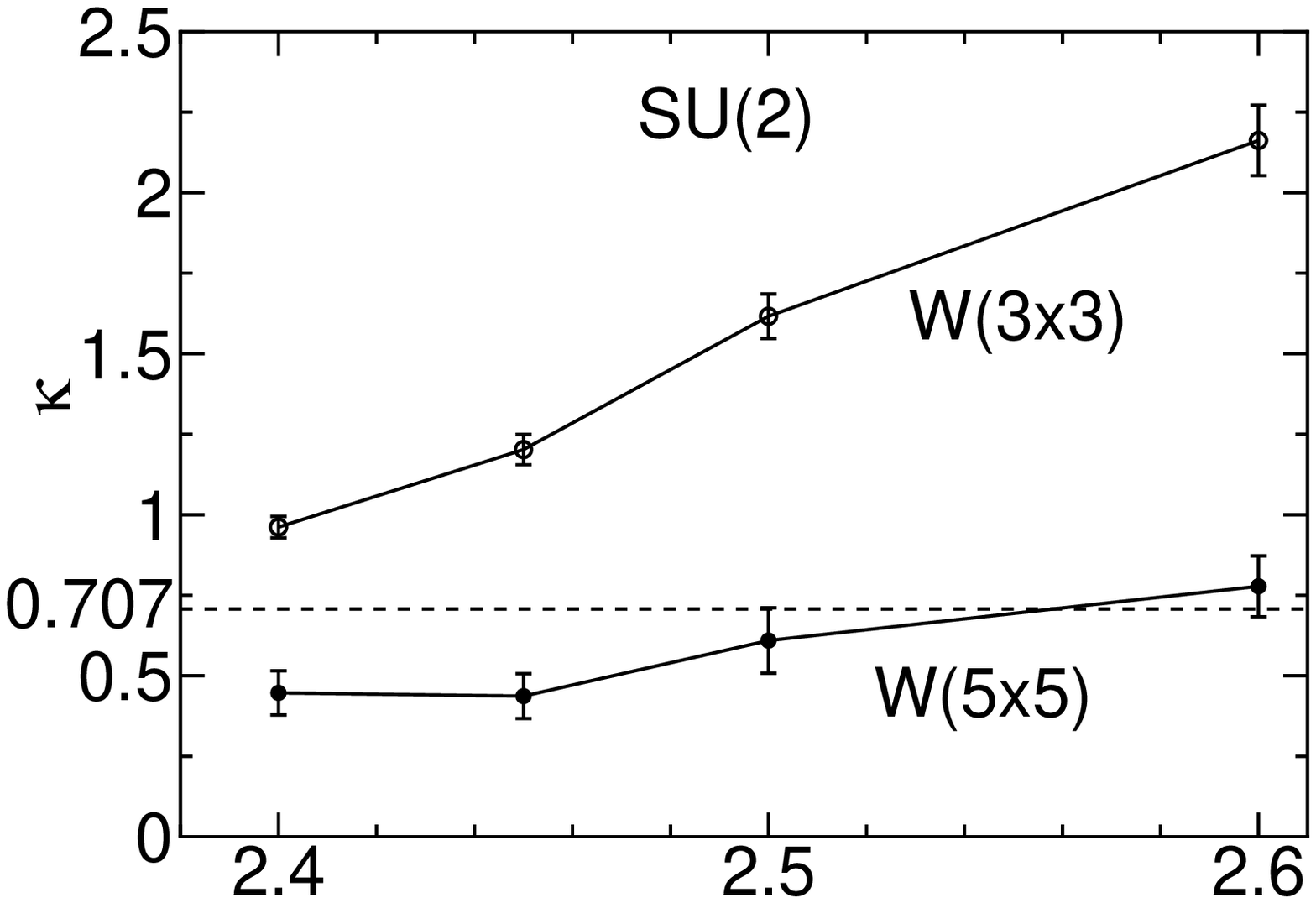}
\end{center}
\vspace{-50mm}
\epsfxsize=0.5\textwidth
\begin{center}
\leavevmode
\epsfbox{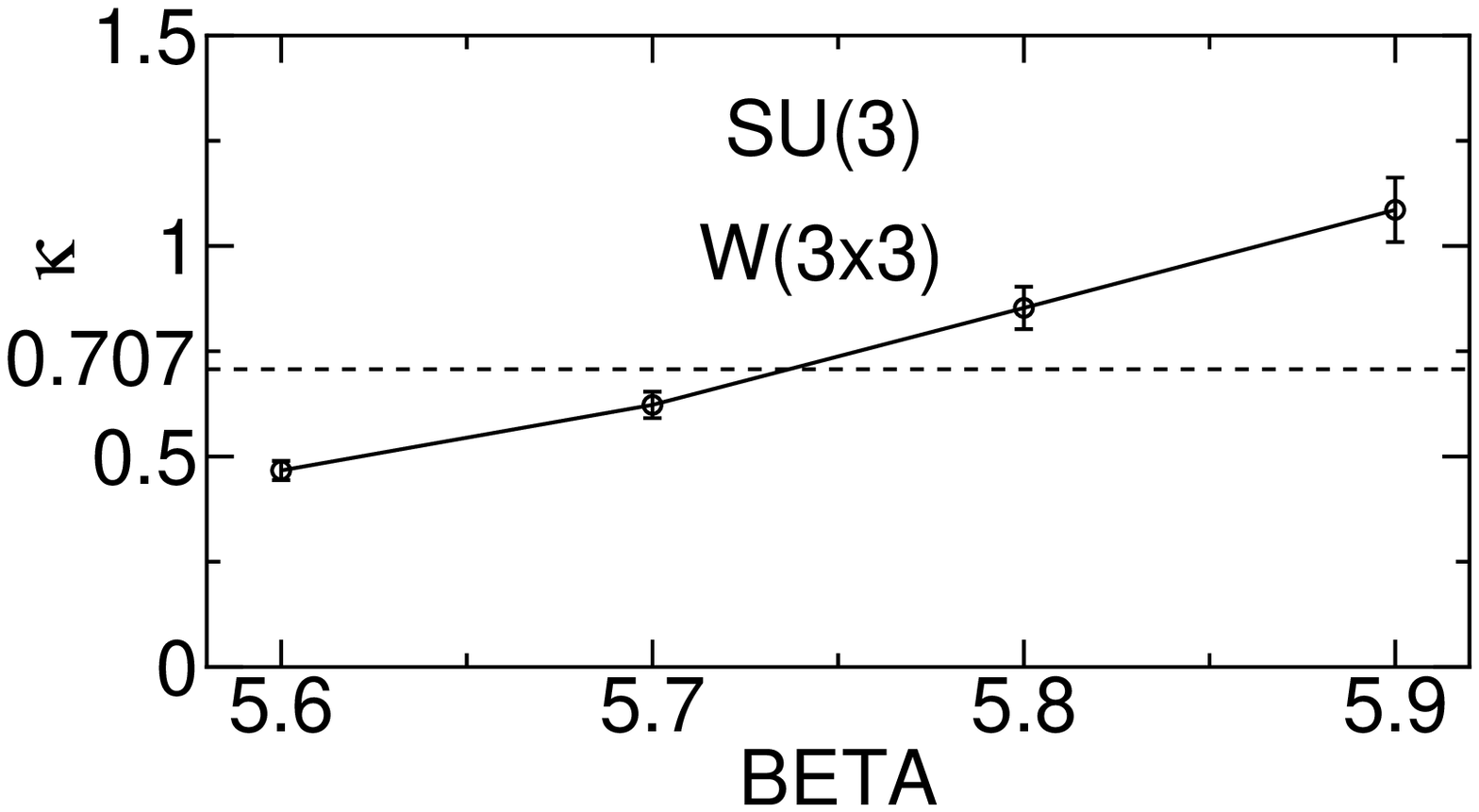}
\end{center}
\vspace{-50mm}
\caption{
The G-L parameter versus $\beta$ in $SU(2)$ and $SU(3)$.
The dotted line denotes the border between the type-1 and the type-2.
}
\vspace{-5mm}
\label{fig:scale3}
\end{figure}

\section{Results}
 We have a direct evidence of the dual Meissner effect 
 in $SU(2)$ and $SU(3)$ QCD.

\begin{enumerate}
\item

The color electric field and the magnetic monopole current satisfy the
extended dual London equation in $SU(2)$ $16^{4}$ and $SU(3)$ $10^{4}$ lattice 
QCD. The monopole current plays the role of the dual "diamagnetic current"
squeezing the electric flux.
\item
All $\lambda$ and $\xi$ data show that the $SU(2)$ and $SU(3)$ vacua are 
   both close to the borderline of the type-1 and type-2 dual 
   superconductor. So, it is expected that the vortex-vortex interactions
   are small. 
\item
It seems that the G-L parameter scales near the borderline in the case of
 the large Wilson loop. Therefore, scalar and axial-vector glueball masses 
 are nearly degenerate:
\begin{eqnarray}
             m(scalar)       =  \sqrt{2}/\xi, \\
             m(axial-vector) =  1/\lambda.
\end{eqnarray}
\item
It is interesting to extend this analysis to the study of 
 the extended monopoles\cite{ivanenko} 
of the type-2, since large monopole loops are seen to be important 
\cite{shiba,suzuk}.
\end{enumerate}

Many thanks to Dr.Singh for giving us detailed information on their analyses.
This work is financially supported by JSPS Grant-in Aid for 
Scientific Research (c)(No.04640289).

\end{document}